\documentclass[showpacs,amsmath,amssymb,aps,showkeys,floatfix,prd,a4paper]{revtex4}

\usepackage[dvips]{graphicx}
\usepackage{dcolumn}
\usepackage{bm}
\usepackage{epsfig}
\usepackage{amsfonts}
\usepackage{amssymb,amscd}

\def\lsim{\raise0.3ex\hbox{$<$\kern-0.75em\raise-1.1ex\hbox{$\sim$}}}

\def\gsim{\raise0.3ex\hbox{$>$\kern-0.75em\raise-1.1ex\hbox{$\sim$}}}

\newcommand{\be}{\begin{equation}}

\newcommand{\ee}{\end{equation}}

\def\beq{\begin{equation}}

\def\eeq{\end{equation}}

\def\beqa{\begin{eqnarray}}

\def\eeqa{\end{eqnarray}}

\newcommand{\ba}{\begin{eqnarray}}

\newcommand{\ea}{\end{eqnarray}}

\def\gappeq{\mathrel{\rlap {\raise.5ex\hbox{$>$}}

{\lower.5ex\hbox{$\sim$}}}}

\def\lappeq{\mathrel{\rlap{\raise.5ex\hbox{$<$}}

{\lower.5ex\hbox{$\sim$}}}}

\def\Toprel#1\over#2{\mathrel{\mathop{#2}\limits^{#1}}}

\begin{document}

\title{True muonium production in  ultraperipheral $PbPb$ collisions}
\author{C. Azevedo$^1$, V.P. Gon\c{c}alves$^1$ and  B.D.  Moreira$^{1,2}$}
\affiliation{$^1$ High and Medium Energy Group, Instituto de F\'{\i}sica e Matem\'atica,  Universidade Federal de Pelotas (UFPel)\\
Caixa Postal 354,  96010-900, Pelotas, RS, Brazil.\\
$^2$ Departamento de F\'isica, Universidade do Estado de Santa Catarina, 89219-710 Joinville, SC, Brazil.  \\
}

\begin{abstract}
In this paper we investigate the production of a true muonium state, which is an atom consisting of a $\mu^+ \mu^-$ bound state, by $\gamma \gamma$ interactions in ultraperipheral $PbPb$ collisions  considering an accurate treatment of the absorptive corrections and for the nuclear form factor. The rapidity  distributions and cross  sections are estimated considering the RHIC, LHC and FCC energies. Our results indicate that the experimental analysis can be useful to observe, for the first time, the true muonium state.
\end{abstract}



\maketitle

\vspace{1cm}

In recent years, the STAR  \cite{star_dilepton},   ALICE \cite{alice_dilepton}, ATLAS \cite{atlas_dilepton} and CMS \cite{cms_dilepton} Collaborations have release data for the dilepton   production by $\gamma \gamma$ interactions in heavy ion collisions at different center -- of -- mass energies and distinct centralities. In particular, these experimental results demonstrated that the equivalent photon approximation \cite{epa} can be applied to describe the ultraperipheral heavy ion collisions (UPHICs), which are characterized by an impact parameter $b$ greater than the sum of the radius of the colliding  nuclei \cite{upc1,upc2,upc3,upc4,upc5,upc6,upc7,upc8,upc9}. In these collisions, the coherent photon -- photon luminosity scales with $Z^4$, where $Z$ is number of protons in the nucleus. As a consequence, such collisions provide an opportunity to study some very rare processes predicted by the quantum electrodynamics (QED). For example, in the last years, the CMS and ATLAS Collaboration have observed, for the first time, the light -- by -- light (LbL) scattering  in ultraperipheral PbPb Collisions   \cite{Aad:2019ock,Sirunyan:2018fhl}. In this case  the elementary  elastic  $\gamma \gamma \rightarrow \gamma \gamma$ process, which  occurs at one -- loop level at order $\alpha^4$ and, consequently, have a tiny cross section, have been enhanced by a large $Z^4$ ($\approx 45 \times 10^6$) factor, becoming feasible for the experimental analysis. Such results strongly motivate the analysis of other final states that can be used to test some of the more important properties of Standard Model (SM). During the last decades, several authors have demonstrated that the study of bound states of dileptons is an ideal testing ground of QED, since it allows to test the properties of leptons,  the charge -- conjugation, parity and time -- reversal (CPT) invariance of the theory as well allows to study the bound state physics (See e.g. Refs. 
  \cite{Stroscio:1975fa, Brodsky:2009gx, Lamm:2013oga, Wiecki:2014ola,Hoyer:2016aew, Bass:2019ibo,Mondal:2019rhs} ). In addition, recent studies have pointed out that such systems provide a probe that is sensitive to beyond SM physics \cite{true_lhcb,true_fixed}. In what follows we will focus on the $\mu^+ \mu^-$ bound state, represented here as $(\mu^+ \mu^-)$ and denoted true muonium (TM) state, which has not been experimentally observed.  In principle, our analysis can be extended for an $e^+ e^-$ bound state: the positronium. However, in this case the Coulomb corrections associated to multiphoton exchange, which are negligible for the TM state, should be taken into account (See discussion in Refs. \cite{serbo_posi,serbo_prc}).

\begin{figure}
\centerline{\psfig{figure=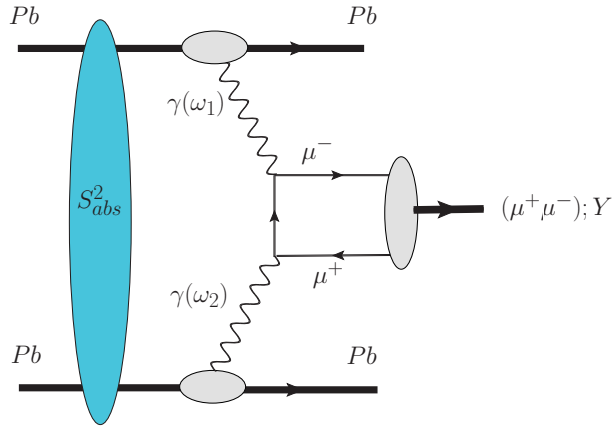,width=8cm}}
\caption{True muonium production by $\gamma \gamma$ interactions in ultraperipheral $PbPb$ collisions.}
\label{fig:diagram}
\end{figure}

The structure of true muonium (TM) is very similar to that of hydrogen. In particular, its ground state can be in a singlet state with spins antipallel and total spin $s=0$, called  para -- TM state, or it can be in the triplet state with spins parallel and total spin $s=1$ denoted ortho -- TM state. However, differently from hydrogen, annihilation can occur in the TM case, with the number of photons $n$ emitted in the decay process being governed by the charge -- conjugation selection rule $(-1)^{l + s} = (-1)^n$, where $l$ is the orbital angular momentum. Consequently, for a true muonium in the ground state, a para-TM state decays into an even number of photons, while the ortho-TM must decay into an odd number.
Therefore, the para and ortho -- TM states can be produced by two  and three photon fusion, respectively. In this paper we focus on the production of para -- TM states in ultraperipheral $PbPb$ collisions.
 We will estimate the cross section and rapidity distribution  using the  equivalent photon approximation, which has been successfully applied for the calculation of the dilepton production in ultraperipheral heavy ion collisions. Our goal is to update and extend the previous estimates presented in the pioneering Refs.  \cite{serbo_jtep, serbo_prc}. Following our previous study of the dilepton production \cite{nos_dilepton}, we will consider a realistic description of the nuclear form factor and for the treatment of the absorptive corrections. We will update the  predictions presented in Ref. \cite{serbo_prc}  for the true muonium production in ultraperipheral $PbPb$ collisions for the RHIC and LHC energies. Moreover, we will  present, for the first time, the predictions for the energies of the  High -- Energy LHC ($\sqrt{s} = 10.6$ TeV) \cite{he_lhc} and Future Circular Collider ($\sqrt{s} = 39$ TeV) \cite{fcc}.  Our study also is motivated by the fact that the resulting  final state is very clean, consisting  of a para--TM state,  two intact nuclei and  two rapidity gaps, i.e. empty regions  in pseudo-rapidity that separate the intact very forward nuclei from the $(\mu^+ \mu^-)$ state. Such aspects can, in principle, be used to separate the events and to probe the TM state.

Initially let's present a brief review of the formalism for the production of a  para -- TM state $(\mu^+ \mu^-)$ in ultraperipheral $PbPb$ collisions, which is represented in Fig. \ref{fig:diagram}. In the equivalent photon approximation \cite{epa}, the associated total cross section can be expressed by \cite{upc2} 
\begin{eqnarray}
\sigma \left(Pb Pb \rightarrow Pb \otimes (\mu^+ \mu^-) \otimes Pb;s \right)   
&=& \int \mbox{d}^{2} {\mathbf b_{1}}
\mbox{d}^{2} {\mathbf b_{2}} 
\mbox{d}W 
\mbox{d}Y \frac{W}{2} \, \hat{\sigma}\left(\gamma \gamma \rightarrow (\mu^+ \mu^-) ; 
W \right )  N\left(\omega_{1},{\mathbf b_{1}}  \right )
 N\left(\omega_{2},{\mathbf b_{2}}  \right ) S^2_{abs}({\mathbf b})  
  \,\,\, .
\label{cross-sec-2}
\end{eqnarray}
where $\sqrt{s}$ is center - of - mass energy of the $PbPb$ collision, $\otimes$ characterizes a rapidity gap in the final state, 
$W = \sqrt{4 \omega_1 \omega_2}$ is the invariant mass of the $\gamma \gamma$ system and
$Y$ is the rapidity of the true muonium in the final state.  The photon energies $\omega_i$ can be expressed in terms of $W$ and $Y$ as follows: 
\begin{eqnarray}
\omega_1 = \frac{W}{2} e^Y \,\,\,\,\mbox{and}\,\,\,\,\omega_2 = \frac{W}{2} e^{-Y} \,\,\,.
\label{ome}
\end{eqnarray}
Moreover, $N(\omega_i, {\mathbf b}_i)$ is the equivalent photon spectrum  
of photons with energy $\omega_i$ at a transverse distance ${\mathbf b}_i$  from the center of nucleus, defined in the plane transverse to the trajectory. The spectrum can be 
expressed in terms of the charge form factor $F(q)$ as follows (See Eq. (2.28) in Ref. \cite{upc2})
\begin{eqnarray}
 N(\omega_i,b_i) = \frac{Z^{2}\alpha_{em}}{\pi^2}\frac{1}{b_i^{2} v^{2}\omega_i}
\cdot \left[
\int u^{2} J_{1}(u) 
F\left(
 \sqrt{\frac{\left( \frac{b_i\omega_i}{\gamma_L}\right)^{2} + u^{2}}{b_i^{2}}}
 \right )
\frac{1}{\left(\frac{b_i\omega_i}{\gamma_L}\right)^{2} + u^{2}} \mbox{d}u
\right]^{2} \,\,.
\label{fluxo}
\end{eqnarray}
In our analysis we will consider the realistic form factor, which corresponds to the Wood - Saxon distribution and is the Fourier transform of the charge density of the nucleus, constrained by the experimental data. It can be analytically expressed by
\begin{eqnarray}
 F(q^{2}) = 
 \frac{4\pi\rho_{0}}{Aq^{3}} 
 \left[ 
 \sin(qR) - qR \cos(qR) 
 \right]
 \left[
 \frac{1}{1 + q^{2} a^{2}}
 \right]
\end{eqnarray}
with $a = 0.549$ fm and $R_{A} = 6.63$ fm \cite{DeJager:1974liz,Bertulani:2001zk}.
 The factor $S^2_{abs}({\mathbf b})$ depends on the impact parameter ${\mathbf b}$ of the $PbPb$ collision and  is denoted the absorptive  factor, which excludes the overlap between the colliding nuclei and allows to take into account only ultraperipheral collisions. Following Ref. \cite{nos_dilepton}, we assume that $S^2_{abs}({\mathbf b})$ can be expressed in terms of the probability of interaction between the nuclei at a given impact parameter, $P_{H}({\mathbf b})$, being given by \cite{Baltz_Klein}
\begin{eqnarray}
 S_{abs}^{2} ({\mathbf b}) 
 =
 1 - P_{H}({\mathbf b}) 
\label{abs2}
 \end{eqnarray}
where
\begin{eqnarray}
 P_{H}({\mathbf b}) = 1 - \exp\left[
 - \sigma_{nn} \int d^{2} {\mathbf r} 
 T_{A}({\mathbf r}) T_{A}({\mathbf r} - {\mathbf b})
 \right].
\end{eqnarray}
with $\sigma_{nn}$ being the total hadronic interaction cross section and $T_{A}$ the nuclear thickness function. Finally, using the Low formula \cite{Low}, we can express the cross section for the production of  the true muonium $(\mu^+ \mu^-)$ 
state due to the two-photon fusion in terms of the two-photon decay width $\Gamma_{(\mu^+ \mu^-) \rightarrow \gamma \gamma}$ as  follows
\begin{eqnarray}
 \hat{\sigma}_{\gamma \gamma \rightarrow (\mu^+ \mu^-)}(\omega_{1},\omega_{2}) = 
8\pi^{2} (2J+1) \frac{\Gamma_{(\mu^+ \mu^-) \rightarrow \gamma \gamma}}{M} 
\delta(4\omega_{1}\omega_{2} - M^{2}) \, ,
\label{Low_cs}
\end{eqnarray}
where $M = 2 m_{\mu}$ and $J$ are, respectively, the mass and spin of the  produced TM state. In the non -- relativistic approximation, one have that only the probability density of $s$ -- states at the origin does not vanish, which implies that $|\Psi_{ns}(0)|^2 = \alpha^3 m_{\mu}^3/8 \pi n^3$. Consequently, we obtain  $\Gamma (n \,^1S_0) = \alpha^5 m_{\mu}/2 n^3$ and that the $\gamma \gamma$ cross section for the lowest TM state is given by
\begin{eqnarray}
 \hat{\sigma}_{\gamma \gamma \rightarrow (\mu^+ \mu^-)}(\omega_{1},\omega_{2}) = 2 \pi^2 \alpha^5 \delta(4\omega_{1}\omega_{2} - M^{2})\,\,.
\end{eqnarray}

\begin{figure}
\centerline{\psfig{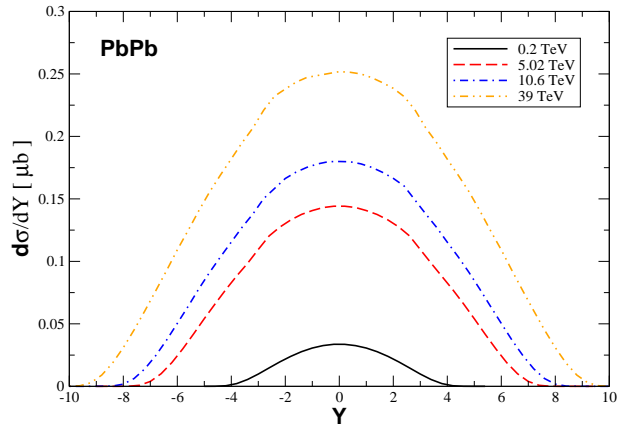}}
\caption{Rapidity distribution for the para -- TM  production by $\gamma \gamma$ interactions in ultraperipheral $PbPb$ collisions.}
\label{fig:rapidity}
\end{figure}

Let's now present our predictions for the rapidity distribution for the production of a para -- TM state with rapidity $Y$ in ultraperipheral $PbPb$ collisions considering different values for the nucleon -- nucleon center -- of -- mass energy. In particular, we will consider the RHIC ($\sqrt{s} = $ 0.2 TeV) and LHC  ($\sqrt{s} = $ 5.5 TeV) energies, as well the proposed energies for the   High -- Energy LHC ($\sqrt{s} = $ 10.6 TeV) and FCC ($\sqrt{s} = $ 39 TeV) \cite{he_lhc,fcc}. In Fig. \ref{fig:rapidity} we present our predictions, which demonstrate that the maximum of distribution occurs for $Y \approx 0$ and that it becomes wider with the increasing of the energy. Moreover, the increase in the value of the distribution for central rapidities from LHC to FCC is of ${\cal{O}} (2)$.
In Table \ref{Tab:cs} we present our predictions for the cross sections considering the full rapidity range covered in $PbPb$ collisions for the different center -- of -- mass energies and two particular range of rapidities, usually covered by central ($-2.5 \le Y \le 2.5$) and forward ($2 \le Y \le 4.5$) detectors. We have that the cross sections increase with the energy and  are larger for central rapidities, in agreement with the results presented in Fig. \ref{fig:rapidity}. Moreover, our results for the LHC energy are similar to those presented in Ref. \cite{serbo_prc}. One important aspect is that our predictions  are of order of  $\mu$b. Assuming the integrated luminosity expected per year for the LHC/HE-LHC/FCC   as being  $350 /\,500  /\,1000  \,$ fb$^{-1}$ \cite{he_lhc},  we predict that the associated  number of events will be larger than  $( 85/\,180/\,500 ) \,\times 10^9$    for these colliders, which  imply that the experimental analysis will be, in principle, feasible. Such large values strongly motivate  a more detailed analysis of the  experimental separation of these events as well to estimate the magnitude of potential backgrounds. In particular, as the dominant decay channel of the para -- TM state will be the decay into two - photons with a small invariant mass, the more important background will be the diphoton system generated in the light -- by -- light (LbL) scattering. Assuming that the associated LbL cross section is known and  constrained by the recent data, such background could be removed, allowing to access the events  associated to the para--TM production. Surely, such aspect deserves a more detailed analysis which we intend to perform in a near future.

\begin{table}[t] 
\centering
\begin{tabular}{||c|c|c|c|c||} 
\hline 
\hline
                        & $\sqrt{s} = $ 0.2 TeV    & $\sqrt{s} = $5.02 TeV            & $\sqrt{s} = $10.6 TeV            & $\sqrt{s} = $39 TeV\\
\hline
\hline
Full rapidity range          &   0.16                           &     1.24        & 1.70                       &    2.74                         \\
\hline 
$-2.5 < Y < 2.5$   &   0.14                           &     0.68        & 0.86                       &    1.22                         \\
\hline 
$2 < Y < 4.5$   &   0.021                           &     0.26        & 0.35                       &    0.52                         \\
\hline
\hline
\end{tabular}
\caption{Cross sections  for the true muonium production in ultraperipheral $PbPb$ collisions considering different rapidity ranges and distinct values of the center -- of -- mass energy. Values  in $\mu$b.} 
\label{Tab:cs}
\end{table}

Finally, let us summarize our main conclusions. In this exploratory study we have investigated the production of the true muonium state in ultraperipheral $PbPb$ collisions at different center -- of -- mass energies. Our main motivation was to estimate the associated cross sections and rapidity distributions in order to verify if this process can be used to observe, for the first time, the QED bound state of a $\mu^+ \mu^-$ pair. Recent studies have demonstrated that the analysis of this state can be useful to test fundamental laws like the CPT theorem as well BSM physics. Motivated by our recent results for the dilepton production \cite{nos_dilepton}, we have used the equivalent photon approximation and considered a realistic model for the nuclear photon flux and for the treatment of the absorptive corrections. 
 We predict large values for the cross sections and event rates, which indicate that a future experimental analysis of the para -- TM state is, in principle, feasible.

\begin{acknowledgments}
VPG acknowledge very useful discussions about photon - induced interactions with Spencer Klein, Mariola Klusek-Gawenda, Daniel Tapia - Takaki and Antoni Szczurek.
This work was  partially financed by the Brazilian funding
agencies CNPq, CAPES,  FAPERGS and INCT-FNA (process number 
464898/2014-5).
\end{acknowledgments}

\hspace{1.0cm}

\end{document}